\documentclass[12pt, draftclsnofoot, onecolumn]{IEEEtran}
\usepackage{epsfig,graphicx,psfrag,amsmath,cases,bm}
\usepackage{latexsym,amssymb,algorithm,mathtools}
\usepackage{algorithmic}
\usepackage{color}
\usepackage{url}
\usepackage{scrtime}
\usepackage{stfloats}
\usepackage{tablefootnote}
\usepackage{cite}
\usepackage{amsfonts,mathabx}
\usepackage{textcomp}
\usepackage{amsthm}
\usepackage{geometry}
\usepackage{enumitem}
\usepackage{titlesec}
\allowdisplaybreaks
\ifCLASSOPTIONcompsoc
    \usepackage[caption=false, font=normalsize, labelfont=sf, textfont=sf]{subfig}
\else
\usepackage[caption=false, font=footnotesize]{subfig}
\fi
\renewcommand{\thesubsubsection}{\arabic{subsubsection}}

\titleformat{\subsubsection}[runin]{\itshape}{\thesubsubsection)}{1em}{}
\titlespacing*{\subsubsection}{\parindent}{0pt}{*1}
\titlespacing*{\section}{0pt}{0.05\baselineskip}{0.05\baselineskip}
\titlespacing*{\subsection}{0pt}{0.05\baselineskip}{0.05\baselineskip}

 \def\BibTeX{{\rm B\kern-.05em{\sc i\kern-.025em b}\kern-.08em
    T\kern-.1667em\lower.7ex\hbox{E}\kern-.125emX}}
    
\usepackage{xcolor,capt-of}
\usepackage{multirow}
\makeatletter
\newcommand*{\rom}[1]{\expandafter\@slowromancap\romannumeral #1@}
\makeatother

\DeclareMathOperator{\mino}{minimize}

\newtheorem{lemma}{Lemma}

\makeatletter
\def\thm@space@setup{\thm@preskip=0pt
\thm@postskip=1pt}
\makeatother

\IEEEaftertitletext{\vspace{-13mm}}
\title{Introducing Resilience to IRS-Assisted Secure Wireless Systems\vspace*{-4mm}}
\author{\vspace*{-0.35cm}\IEEEauthorblockN {Yifei Wu\IEEEauthorrefmark{1}, Mehmet Emin Arslan\IEEEauthorrefmark{2}, Niels Neumann\IEEEauthorrefmark{2}, Wolfgang Gerstacker\IEEEauthorrefmark{1}, and Robert Schober\IEEEauthorrefmark{1}}\\
\IEEEauthorrefmark {1}Friedrich-Alexander-Universit\"at
Erlangen-N\"urnberg, Germany;
\IEEEauthorrefmark {2}Technische Universit\"at Clausthal, Germany\\[-10pt]\vspace*{-0.5cm}}
\begin{document}
\maketitle
\begin{abstract}
Intelligent reflecting surfaces (IRSs) are envisioned as a transformative method for enhancing the physical layer security of future communication networks. However, current IRS-assisted secure wireless system designs incur high time cost for the joint optimization of beamforming vectors and IRS phase shifts and are susceptible to unforeseen disruption. Therefore, introducing resilience in IRS-assisted secure systems is essential. In this paper, we first quantify the resilience performance as the combination of the absorption and adaptation performance. In particular, the absorption performance measures the system's robustness when facing failure while the adaptation performance reflects the system's ability to recover from failure. Then, we propose a two-timescale transmission protocol aiming to enhance system resilience while limiting maximum information leakage to eavesdroppers. Specifically, in the initialization phase, the base station (BS) beamforming vectors and IRS phase shifts are optimized iteratively using an alternating optimization (AO) approach to improve the absorption performance of the system, i.e., the worst-case achievable rate, given a predefined maximum tolerable leakage rate, by accounting for the impact of long-term channel variations. When the system detects an outage, a fast system recovery mechanism is activated to restore operation by adjusting only the BS beamforming vectors, which enhances the adaptation performance of the system. Our numerical results reveal that the proposed algorithm significantly improves system resilience, including system robustness and recovery performance. Moreover, the proposed design guarantees outage-free transmission over a long time interval for moderate-size IRSs.
\end{abstract}
\vspace{-0.5mm}\section{Introduction}\vspace{-0.2mm}
\footnotetext{\hspace{-0mm}This work has been funded by the Federal
Office for Information Security (BSI) within the project RIS4NGWB under grant 01MO23001A-C.}In recent years, physical layer security (PLS) has received considerable attention from both academia and industry. 
Specifically, PLS leverages the inherent randomness of wireless channels to restrict the information accessible by potential eavesdroppers (Eves), entailing low computational complexity and providing high efficiency. Various approaches for enhancing PLS in wireless networks have been proposed in the literature, including cooperative relaying and cooperative jamming of Eves \cite{yu2020robust}. However, these methods employ active relays and additional helpers for security enhancement, leading to high hardware costs and high energy consumption. Moreover, due to the randomness of the wireless channel, it is challenging to ensure satisfactory secrecy performance, even if jamming signals are injected\cite{hu2021robust}. 
To address these shortcomings, intelligent reflecting surfaces (IRSs) have been advocated as a transformative technique for PLS. Comprising cost-effective passive and programmable elements, IRSs possess the capability to intelligently establish favorable wireless propagation environments. By leveraging this appealing property, one can adjust the IRS phase shifts such that the signal reflected from the IRS strengthens the received signal at the legitimate users while destructively adding to the signals received by Eves. Motivated by these benefits, numerous works have explored the application of IRSs to improve PLS performance \cite{yu2020robust,asaad2022secure,hu2021robust}. However, while these works improve secrecy by mitigating signal leakage and interference, they remain susceptible to unforeseen disruptions, including physical blockages and environmental interference. Moreover, adjusting the IRS phase shifts in each time frame introduces high signal processing complexity and large overhead between base station (BS) and IRS.

Addressing these challenges requires the introduction of resilience strategies that maintain consistent system performance and security, even under adverse conditions. Resilience is the capacity of a system to absorb disturbances and to reorganize and adapt to changes while preserving its essential functions \cite{weinberger2023ris,najarian2019design}. In practical terms, resilience reflects a system’s ability to maintain functionality when errors occur, adapt to disruptive impairments, and recover functionality swiftly.
To achieve this, the system must continuously monitor network conditions and be able to apply corrective measures that prevent or mitigate potential outages, thus offering a robust response to failures. This concept is especially crucial in secure wireless communication systems since security performance is very sensitive to environmental changes and channel estimation errors\cite{hu2021robust,yu2021robust}. Moreover, resilience can be characterized into absorption and adaptation performance, which reflect a system's ability to maintain and recover functionality when facing errors, respectively \cite{weinberger2023ris, najarian2019design}. In the context of communication systems, the absorption performance can be quantified by the robust achievable rate under long-term channel variations. However, due to unforeseen channel variations, system outages may still occur during transmission. Then, the adaptation performance measures the system's achievable rate after the system has taken action to recover from a system outage. In this paper, to enhance the resilience of the system under consideration, we propose a two-timescale joint BS and IRS beamforming design. Specifically, after the long-term design establishes the initial robust BS beamforming vectors and IRS configuration, the system continuously monitors network performance. If any indications of system outage arise, a short-term resilience mechanism is triggered to adjust the BS beamforming vectors and counteract the outage, ensuring sustained functionality.
The main contributions of this paper can be summarized as follows:\\[-15pt]
\begin{itemize}
\item To facilitate the optimization of the system's resilience, we first define and model the resilience performance for communication systems based on \cite{weinberger2023ris, najarian2019design}. In particular, the absorption and adaptation performances are modeled as the worst-case achievable rate under channel variations and the achievable rate after system recovery, respectively. 
    \item To achieve both system security and resilience, we propose a novel two-timescale joint beamforming scheme for an IRS-assisted multiuser multiple-input single-output (MISO) system.
    Then, we propose a series of mathematical transformations that allow us to recast the considered long-term and short-term problems into tractable mixed integer nonlinear programming (MINLP) and fractional programming (FP) problems with several linear matrix inequality (LMI) constraints, respectively.
\item We propose two iterative algorithms based on Lagrangian dual and quadratic transformations \cite{shen2018fractional} to obtain suboptimal solutions of the considered long-term and short-term design problems, respectively. Our numerical results reveal that the proposed design significantly improves the system's resilience, including the absorption and recovery performance.\vspace{-0.2cm}
\end{itemize}
\textit{Notation:} 
Vectors and matrices are denoted by boldface lower case and boldface capital letters, respectively. $\mathbb{R}^{N\times M}$ and $\mathbb{C}^{N\times M}$ represent the space of $N\times M$ real-valued and complex-valued matrices, respectively. $|\cdot|$ and $||\cdot||_2$ stand for the absolute value of a complex scalar and the $l_2$-norm of a vector, respectively. $(\cdot)^T$, $(\cdot)^*$, and $(\cdot)^H$ denote the transpose, the conjugate, and the conjugate transpose of their arguments, respectively. $\mathrm{Tr}(\cdot)$ is the trace of the input argument. $\mathbf{0}_{L}$ represents the all-zeros column vector of length $L$. $\mathbf{A}\succeq\mathbf{0}$ indicates that $\mathbf{A}$ is a positive semidefinite matrix. $\mathrm{diag}(\mathbf{a})$ denotes a diagonal matrix whose main diagonal elements are given by the entries of vector $\mathbf{a}$. $\bigotimes$ stands for the Kronecker product between two matrices. 
\section{System Model}
\subsection{Signal Model}
\begin{figure}\vspace{-2mm}
    \centering
    \includegraphics[width=2.0in]{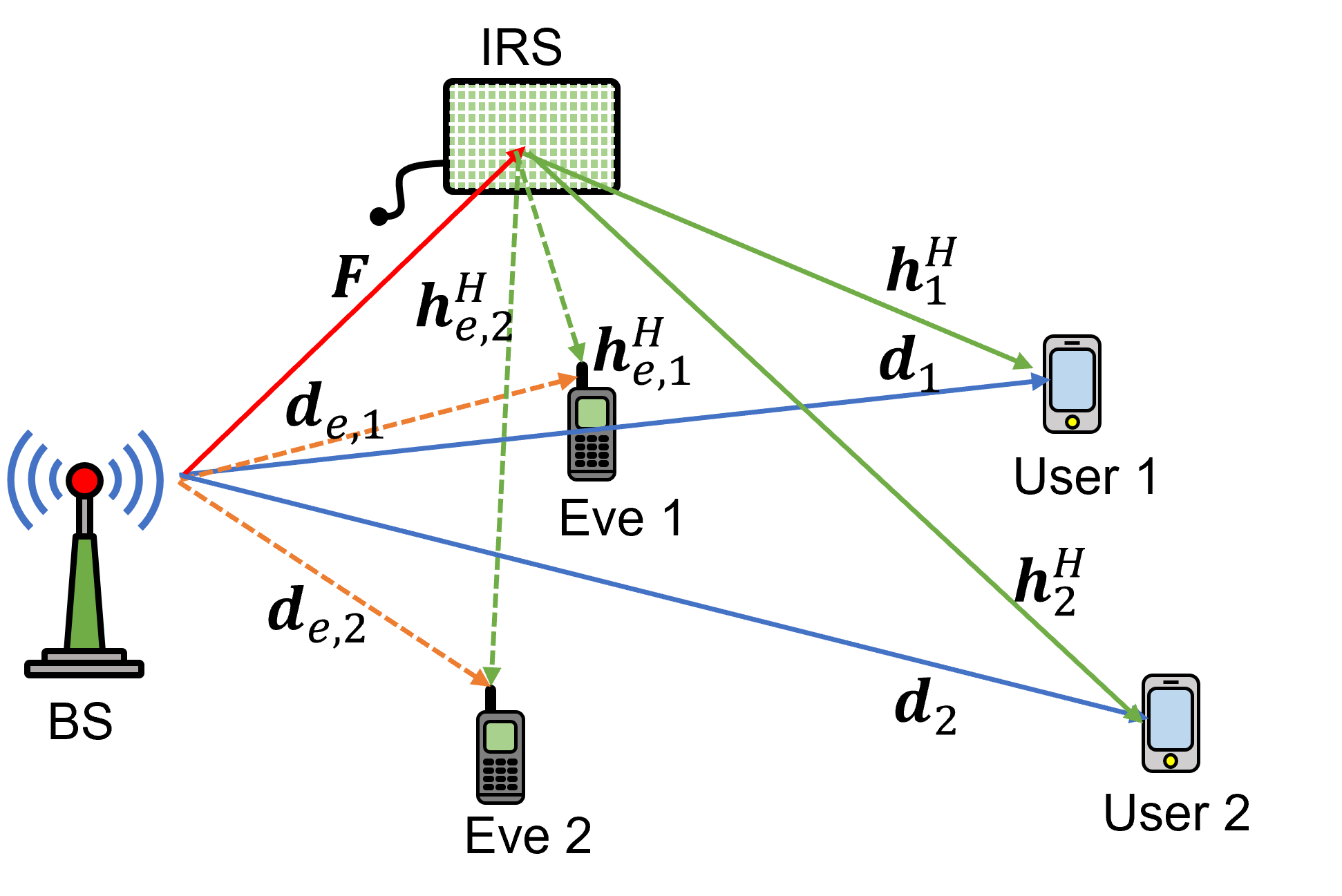}
    \caption{Considered IRS-assisted system with $K=2$ users and $J=2$ Eves.}
    \label{fig:enter-label}
\end{figure}
\begin{figure}\vspace{-7mm}
    \centering
    \includegraphics[width=3.6 in]{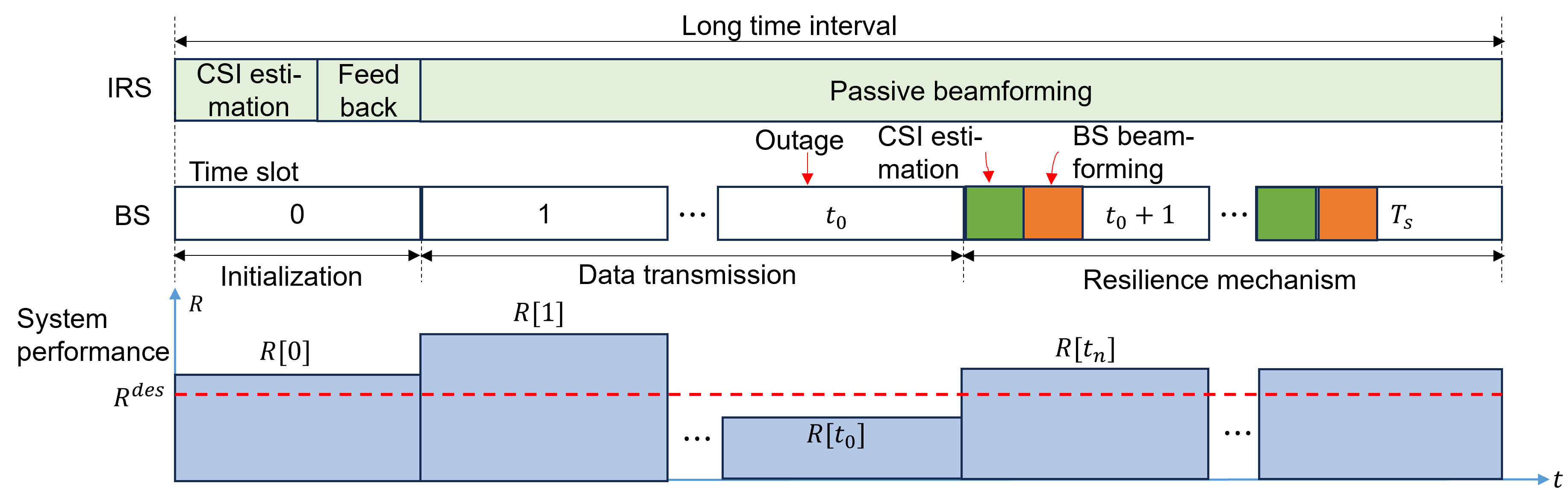}
    \caption{Frame structure of the proposed transmission protocol.}
    \label{fig:enter-label}
\end{figure}
 As depicted in Fig. 1, we consider a multiuser downlink communication system consisting of a BS, $K$ users, and $J$ Eves equipped with single omnidirectional antenna. The BS is equipped with $M$ antenna elements and serves $K$ single-antenna users. 
 Moreover. an IRS comprising $N$ phase shift elements is deployed in the network. Adjusting the IRS phase shifts in each time slot requires frequent communication between BS and IRS, resulting in extra time and overhead costs \cite{zhao2020intelligent}, making real-time IRS tuning impractical. Therefore, we consider a long-term joint design of the BS beamforming vectors and the IRS phase shift matrix, as shown in Fig. 2. Due to potential changes in the environment (e.g., blockages, user movement), the channels are not constant over the considered long time frame $T_{\mathrm{LT}}$. To address this issue, we divide the considered long time frame into $T_s$ short time slots, during which the channels are assumed to be constant.  Then, the received signals of user $k$, ${y}_{k}[t_s]$, and Eve $j$, ${y}_{e,j}[t_s]$, in time slot $t_s$ are given as
\begin{equation}
    \label{orichannel1}
    \begin{aligned}
    {y}_{k}[t_s]&=\left(\mathbf{h}_{k}^H[t_s]\bm{\Phi}\mathbf{F}[t_s]+\mathbf{d}^H_{k}[t_s]\right)\sum_{k\in\mathcal{K}}\mathbf{w}_k s_k+{n}_k,\\[-8pt]
    &\overset{(a)}{=}\mathbf{v}^H{\mathbf{H}}_k[t_s]\hat{\mathbf{F}}_k[t_s]\sum_{k\in\mathcal{K}}\mathbf{w}_k s_k+{n}_k,
    \end{aligned}
\end{equation}
\vspace{1mm}\begin{equation}
    \label{oriChannel_Eve1} 
    \begin{aligned}
        \hspace{-2mm}{y}_{e,j}[t_s]&=\left(\mathbf{h}^H_{e,j}[t_s]\bm{\Phi}\mathbf{F}[t_s]+\mathbf{d}^H_{e,j}[t_s]\right)\sum_{k\in\mathcal{K}}\mathbf{w}_k s_k+{n}_{e,j}\\[-8pt]
        &\overset{(b)}{=}\mathbf{v}^H{\mathbf{H}}_{e,j}[t_s]\hat{\mathbf{F}}_{e,j}[t_s]\sum_{k\in\mathcal{K}}\mathbf{w}_k s_k+{n}_{e,j}.
    \end{aligned}
\end{equation}
respectively, where $\mathbf{h}_{k}[t_s]\in\mathbb{C}^{N}$, $\mathbf{F}[t_s]\in\mathbb{C}^{N\times M}$. and $\mathbf{d}_{k}[t_s]\in\mathbb{C}^{M}$ denote the conjugate channel between IRS and user $k$, the channel matrix between BS and IRS, and the conjugate channel between BS and user $k$ in time slot $t_s$, respectively. Besides, the conjugate channel between IRS and Eve $j$ and the conjugate channel between BS and Eve $j$ in time slot $t_s$ are denoted as $\mathbf{h}_{e,j}[t_s]\in\mathbb{C}^{N}$ and $\mathbf{d}_{e,j}[t_s]\in\mathbb{C}^{M}$, respectively. Moreover, $\mathbf{w}_k\in \mathbb{C}^{M\times 1}$ represents the linear beamforming vector for user $k$, and  $\boldsymbol{\Phi}=\operatorname{diag}\left(e^{j\theta_1}.\cdots,e^{j\theta_N}\right)$ is the phase shift matrix of the IRS for the considered long time frame. Moreover, to obtain reformulations (a) and (b), we introduce $\hat{\mathbf{F}}_k[t_s]=[\mathbf{F}^H[t_s],\mathbf{d}_k[t_s]]^H$, $\hat{\mathbf{F}}_{e,j}[t_s]=[\mathbf{F}^H[t_s],\mathbf{d}_{e,j}[t_s]]^H$, ${\mathbf{H}}_k[t_s]=\mathrm{diag}([{\mathbf{h}}_k^H[t_s],1]),$ ${\mathbf{H}}_{e,j}[t_s]=\mathrm{diag}([{\mathbf{h}}_{e,j}^H[t_s],1]),$ and $\mathbf{v}=[e^{j\theta_1},\cdots,e^{j\theta_N},1]^H$.  In this paper, we consider the practical case, where each element of the IRS can assume only $L$ different discrete phase shift values, i.e.,  $\theta_n\in\bm{\Psi}=\{\Psi_1,\cdots,\Psi_L\}$.
Moreover, $s_k\in\mathbb{C}$ denotes the symbol transmitted to user $k$ and $\mathbb{E}[|s_k|^2]=1$. Furthermore, $n_k\sim \mathcal{CN}(0,\sigma_k^2)$ and $n_{e,j}\sim \mathcal{CN}(0,\sigma_{e,j}^2)$ denote the noise at user $k$ and Eve $j$, respectively. For convenience, we define sets $\mathcal{M}=\{1,\cdots,M\}$, $\mathcal{K}=\{1,\cdots,K\}$, $\mathcal{N}=\{1,\cdots,N\}$, and $\mathcal{T}_s=\{1,\cdots,T_s\}$ to collect the indices of the BS transmit antennas, users, IRS elements, and time slots, respectively. Also, we assume that the BS has maximum transmit power $P_{\mathrm{max}}$, i.e., $ \sum_{k\in\mathcal{K}}\|\mathbf{w}_k\|^2_2\leq P_{\mathrm{max}}.   $
Based on the signal model in \eqref{orichannel1}, the received signal-to-interference-plus-noise ratio (SINR) at user $k$ in time slot $t_s$ is given by
\begin{equation}
\begin{aligned}
    \hspace*{-3mm}\mathrm{SINR}_k[t_s]
    &=\frac{\mathrm{Tr}(\widetilde{\mathbf{H}}_k^H[t_s]\mathbf{V}\widetilde{\mathbf{H}}_k[t_s]\mathbf{W}_k)}{\sum_{k'\in\mathcal{K}\setminus k}\mathrm{Tr}(\widetilde{\mathbf{H}}_k^H[t_s]\mathbf{V}\widetilde{\mathbf{H}}_k[t_s]\mathbf{W}_{k'})+\sigma_k^2},
\end{aligned}
\end{equation}
where $\mathbf{V}=\mathbf{v}\mathbf{v}^H$, $\mathbf{W}_k=\mathbf{w}_k\mathbf{w}_k^H$, and $\widetilde{\mathbf{H}}_k[t_s]={\mathbf{H}}_k[t_s]\hat{\mathbf{F}}_k[t_s]$ denotes the overall channel between BS and user $k$ in time slot $t_s$. Then, the achievable rate of user $k$ in time slot $t_s$ is given by $R_k[t_s]=\log(1+\mathrm{SINR}_k[t_s])$. Next, for secrecy provisioning, we adopt the worst-case secrecy assumption that the potential Eves have unlimited computational resources to cancel all interference before decoding the desired information. Thus, the channel capacity of Eve $j$ for decoding the signal of user $k$ in time slot $t_s$ is given by
\begin{equation}
 \begin{aligned}
\widehat{R}_{k,j}[t_s]=\log\big(1+\frac{1}{\sigma_{e,j}^2}\mathrm{Tr}(\widetilde{\mathbf{H}}_{e,j}^H\mathbf{V}\widetilde{\mathbf{H}}_{e,j}[t_s]\mathbf{W}_k[t_s])\big),
 \end{aligned}   
\end{equation}
where $\widetilde{\mathbf{H}}_{e,j}[t_s]={\mathbf{H}}_{e,j}[t_s]\hat{\mathbf{F}}_{e,j}[t_s]$ denotes the overall channel between BS and Eve $j$ in time slot $t_s$. To guarantee the secrecy of the considered system, we demand that the maximum information leakage to Eve is below a desired level, i.e.,
\begin{equation}\label{Secure_constraint}
    \widehat{R}_{k,j}[t_s]\leq R^S_{k,j},\forall k,j,t_s,
\end{equation}
where $R^S_{k,j}$ is a predefined parameter which stands for the maximum tolerable information leakage to Eve $j$ for wiretapping the signal transmitted to user $k$.
\subsection{CSI Acquisition and Resilience Strategy}
Fig. 2 illustrates the proposed transmission protocol. In the initialization time slot, the BS estimates the channel state information (CSI) of all legitimate channels. To this end, the users send pilot signals to the BS through the IRS to enable channel estimation. Thus, the BS is able to obtain the accurate CSI of the users for the initialization time slot. Moreover, we assume that the channel variations in a given time interval are upper-bounded. In particular, we model the channel variation between the overall channel of user $k$ in the first time slot, $\widetilde{\mathbf{H}}_k[0]$, and the overall channel of user $k$ in time slot $t_s$, $\widetilde{\mathbf{H}}_k[t_s]$, by a norm-bounded model, which is given by
\begin{equation}
        \begin{aligned}
        \widetilde{\mathbf{H}}_{k}[t_s]&=\widetilde{\mathbf{H}}_k[0]+\Delta\widetilde{\mathbf{H}}_k[t_s],\quad \|\Delta\widetilde{\mathbf{H}}_k[t_s]\|_F\leq \epsilon_{ \widetilde{\mathbf{H}}_{k}}, 
    \end{aligned}
\end{equation}
where $\Delta\widetilde{\mathbf{H}}_k[t_s]$ denotes the possible CSI variation of user $k$ between the initialization time slot and time slot $t_s$\footnote{For simplicity, we assume that the $\Delta\widetilde{\mathbf{H}}_k[t_s]$ in all time slots follow the same norm-bounded distribution. In future work, we will consider a more practical model for $\epsilon_{ \widetilde{\mathbf{H}}_{k}}$.}. Note that given the proposed transmission protocol, $\widetilde{\mathbf{H}}_k[0]$ is assumed to be perfectly known at the BS in time slot $0$. 

Moreover, since Eves typically attempt to conceal their presence from the BS, they may not cooperate for CSI acquisition. Thus, the BS utilizes the signal leakage from the Eves for channel estimation, leading to imperfect CSI knowledge, i.e., 
\begin{equation}\label{Eve_CSI_uncertainty}
        \begin{aligned}
        \hspace{-3mm}\widetilde{\mathbf{H}}_{e,j}[t_s]&=\widehat{\widebar{\mathbf{H}}}_{e,j}+\Delta\widetilde{\mathbf{H}}_{e,j}[t_s],\, \|\Delta\widetilde{\mathbf{H}}_{e,j}[t_s]\|_2\leq \epsilon_{\widetilde{\mathbf{H}}_{e,j}},
    \end{aligned}
\end{equation}
where $\widehat{\widebar{\mathbf{H}}}_{e,j}$ denotes the estimate of the overall channel between the BS and Eve $j$. Moreover, $\Delta\widetilde{\mathbf{H}}_{e,j}[t_s]$ and $\epsilon_{\widetilde{\mathbf{H}}_{e,j}}$ represent the CSI deviation between the channel estimate and the channel $\widetilde{\mathbf{H}}_{e,j}[t_s]={\mathbf{H}}_{e,j}[t_s]\hat{\mathbf{F}}_{e,j}[t_s]$ in time slot $t_s$ of Eve $j$.

Based on the measured CSI, the BS jointly designs the BS beamforming matrix and IRS phase shifts to guarantee the desired rates of all users while satisfying the security constraints in \eqref{Secure_constraint}. Then, we fix the BS beamforming vectors and IRS phase shifts for the following time slots as long as no system failure occurs. To detect system failure, users are required to report their achievable rate to the BS. If the achievable rate of user $k$ is below the desired rate in time slot $t_0$, a system failure is reported to the BS\footnote{Since the Eves typically attempt to hide their existence from the BS, monitoring the achievable rate of Eves is not feasible for system design. Therefore, we limit the capacity of Eves to a given threshold for all time slots with the CSI uncertainty model in \eqref{Eve_CSI_uncertainty}.}, which then activates the resilience protocol to restore system performance by re-estimating the effective channel vectors $\widehat{\mathbf{h}}_k[t_0]=\widetilde{\mathbf{H}}^H_k[t_0]\mathbf{v}$ between the BS and all users, which is low dimensional and can be accurately estimated at the BS \cite{zhao2020intelligent}. Thus, the perfect CSI of the effective channel vectors $\widehat{\mathbf{h}}_k[t_0]$, $\forall k$, is assumed to be available at the BS for the resilience mechanism design. Then, the beamforming vectors are adjusted based on the re-estimated effective channel vectors.

\subsection{Resilience Metrics}
In this subsection, we introduce performance metrics for resilience. 
For the considered system, we define a system failure as the inability to meet the rate requirement of any user. As shown in Fig. 2, there are two crucial time instants for resilience, i.e., the initial time slot when the system failure occurs, $t_0$, and the time slot of recovery, $t_n$.
Based on the resilience function proposed in \cite{weinberger2023ris,najarian2019design}, we define the system absorption and adaptation metrics as follows
\begin{equation}
    r_{\mathrm{abs}}=\frac{1}{K}\sum_{k\in\mathcal{K}}\frac{R_k[t_0]}{R_k^{\mathrm{des}}},\,\,r_{\mathrm{ada}}=\frac{1}{K}\sum_{k\in\mathcal{K}}\frac{R_k[t_n]}{R_k^{\mathrm{des}}},\notag
\end{equation}
where $R_k[t_0]$ and $R_k[t_n]$ denote the achievable rate of user $k$ after system failure and recovery, respectively. Moreover, $R_k^{\mathrm{des}}$ denotes the desired rate of user $k$. Note that for an outage-free system, i.e., $R_k[t_0]\geq R_k^{\mathrm{des}}$, we have for the absorption metric $r_{\mathrm{abs}}\geq 1$. Moreover, if a system is able to fully recover from system failure, i.e., $R_k[t_n]\geq R_k^{\mathrm{des}}$, we can obtain for the adaption metric $r_{\mathrm{abs}}\geq 1$. 
Here, system absorption measures robustness, which can be induced by anticipatory actions in the initialization phase, while system adaptation reflects the recovery performance induced by appropriate actions post failure.

\section{Problem Formulation}



\subsection{Initialization Problem}
First, we investigate the system design for the initialization phase to optimize the absorption performance. In this work, we aim to guarantee that the system maintains a desired level of performance after system failure. Thus, we model the absorption performance as the worst-case absorption metric after failure, which is given by
\begin{equation}
    \hspace{-1mm}r_{\mathrm{abs}}^{\mathrm{worst}}\hspace*{-0.5mm}=\hspace*{-0.5mm}\frac{1}{KT_s}\sum_{\substack{t_0\in\mathcal{T}_s,\\k\in\mathcal{K}}}\underset{\widetilde{\mathbf{H}}_k[t_0]}{\min}\frac{R_k[t_0]}{R_k^{\mathrm{des}}}\hspace*{-0.5mm}\overset{(a)}{=}\hspace*{-0.5mm}\frac{1}{K}\sum_{k\in\mathcal{K}}\underset{\widetilde{\mathbf{H}}_k[t_0]}{\min}\frac{R_k[t_0]}{R_k^{\mathrm{des}}},
\end{equation}
Here, the worst-case absorption metric is adopted for resilient system design since a resilient system should be able to absorb unexpected disruptions caused by channel variations\cite{najarian2019design}. Moreover, since the exact time of system failure is unknown in the initialization phase, we aim to enhance the average absorption performance across all time slots. Moreover, the equality (a) is attributed to the fact that the overall channel $\Delta\widetilde{\mathbf{H}}_k[t_s]$, $\forall t_s$ follows the same norm-bounded error model across all time slots $t_0\in\mathcal{T}_s$.
Here, in the initialization phase, we aim to optimize the absorption metric $r_{\mathrm{abs}}^{\mathrm{worst}}$ while limiting information leakage to each Eve. Then, the absorption optimization problem for the initialization design is formulated as follows,
\begin{eqnarray}
\label{Ori_problem}
     &&\hspace*{-6mm}\underset{\mathbf{v},\mathbf{W}_k}{\mino}\hspace*{2mm} -r_{\mathrm{abs}}^{\mathrm{worst}}\notag\\[-3pt]
    &&\hspace*{-5mm}\mbox{s.t.}\hspace*{2mm}\hspace*{0mm} \mbox{C1:}\hspace*{1mm}\theta_n\in\bm{\Psi},\hspace*{1mm}\forall n,\mbox{C2:}\hspace*{1mm}\sum_{k\in\mathcal{K}}\mathrm{Tr}(\mathbf{W}_k)\leq P_{\mathrm{max}}\notag\\[-5pt]
    &&\hspace*{1mm}\mbox{C3:}\hspace*{1mm}\max_{\widetilde{\mathbf{H}}_{e,j}[t_s]} \hat{R}_{k,j}[t_s]\leq R^S_{k,j}, \forall j,k,t_s\\[-3pt]
    &&\hspace*{1mm}\mbox{C4:}\hspace*{1mm}\mathbf{V}=\mathbf{v}\mathbf{v}^H,\hspace*{1mm} \mbox{C5:}\hspace*{1mm}\mathbf{W}_k\succeq \mathbf{0}, \forall k, \hspace*{1mm} \mbox{C6:}\hspace*{1mm} \mathrm{rank}(\mathbf{W}_k)=1\hspace*{1mm},\notag\
\end{eqnarray}
where constraints C1 and C2 are imposed because of the discrete nature of the IRS phase shifts and the maximum power limitation of the BS, respectively. Secrecy constraint C3 aims to limit the worst-case information leakage to all Eves below a given threshold $R^S_{k,j}$. Note that the formulated problem is non-convex due to the non-convexity of the objective function, the coupling between $\mathbf{W}_k$ and $\mathbf{v}$ and the discrete nature of $\theta_n$. In Section \rom{4}, we introduce a series of transformations to recast problem \eqref{Ori_problem} into a tractable form and propose an AO-based method to obtain a suboptimal solution.

\subsection{Recovery Problem}
Here, we aim to maximize the adaptation performance under power and security constraints. Since the exact time of system recovery is unknown, we maximize the system performance for each time slot after system failure. Here, a system failure occurs if the achievable rate of any user falls below the desired rate. Moreover, since the Eves may not report their achievable rates to the BS, we do not account for failures due to Eves. Instead, we limit the maximum information leakage to Eves below a desired level in the recovery phase. Then, in the time slots $t_s$ following a system failure, the BS first estimates the effective channels $\widetilde{\mathbf{h}}_k[t_s]$, $\forall k$ between the BS and the users. Since the effective channel $\widetilde{\mathbf{h}}_k[t_s]$ has a lower dimension compared to the overall channel $\widetilde{\mathbf{H}}_k[t_s]$, accurate and fast channel estimation is feasible. 
Thus, the resource allocation problem for the recovery phase is formulated as follows:
\begin{eqnarray}
\label{Ori_problem_Ada}
     &&\hspace*{-6mm}\underset{\mathbf{W}_k}{\mino}\hspace*{2mm} -\frac{1}{K}\sum_{k\in\mathcal{K}}\frac{\log\left(1+\overline{\mathrm{SINR}}_k[t_s]\right)}{R_k^{\mathrm{des}}}\notag\\[-3pt]
    &&\hspace*{-1mm}\mbox{s.t.}\hspace*{5mm} \mbox{C2},\mbox{C3},\mbox{C5},\mbox{C6},
\end{eqnarray}
where $\overline{\mathrm{SINR}}_k[t_s]$ denotes the received SINR at user $k$ in time slot $t_s$, which is given by
\begin{equation}
\overline{\mathrm{SINR}}_k[t_n]=\frac{\mathrm{Tr}(\widebar{\mathbf{H}}_k[t_s]\mathbf{W}_k)}{\sum_{k'\neq k}\mathrm{Tr}(\widebar{\mathbf{H}}_k[t_s]\mathbf{W}_{k'})+\sigma_k^2},   
\end{equation}
with $\widebar{\mathbf{H}}_k[t_s]=\widetilde{\mathbf{h}}_k[t_s]\widetilde{\mathbf{h}}_k^H[t_s]$.
In Section \rom{4}, we exploit the FP transformation in \cite{shen2018fractional} to recast \eqref{Ori_problem_Ada} to a tractable form and propose an AO-based method to obtain a suboptimal solution.
\section{Algorithm Design}
\subsection{Absorption Optimization Design}
\subsubsection{Problem Reformulation: } 
First, based on the monotonically increasing nature of the $\log(\cdot)$ function, we derive a lower bound on the objective function in \eqref{Ori_problem} by introducing two slack variable vectors $\bm{\eta}=[\eta_1,\cdots, \eta_K]^H$ and $\bm{\zeta}=[\zeta_1,\cdots, \zeta_K]^H$ as follows:
\begin{equation}
    \widebar{r}_{\mathrm{abs}}(\bm{\eta},\bm{\zeta})=\frac{1}{K}\sum_{k\in\mathcal{K}}\frac{1}{R_k^{\mathrm{des}}}\log\left(1+\frac{\eta_k}{\zeta_k+\sigma_k^2}\right),
\end{equation}
where the following two inequality constraints are used 
\begin{equation}
\mbox{C7:}\,\eta_k\leq \min_{\widetilde{\mathbf{H}}_k[t_0]}\mathrm{Tr}(\widetilde{\mathbf{H}}_k^H[t_0]\mathbf{V}\widetilde{\mathbf{H}}_k[t_0]\mathbf{W}_k)
\end{equation}
\begin{equation}
    \mbox{C8:}\, \zeta_k\geq \max_{\widetilde{\mathbf{H}}_k[t_0]}\sum_{k'\in\mathcal{K}\setminus k}\mathrm{Tr}(\widetilde{\mathbf{H}}_k^H[t_0]\mathbf{V}\widetilde{\mathbf{H}}_k[t_0]\mathbf{W}_{k'}).
\end{equation}
Then, by applying the Lagrangian dual transformation from \cite[Theorem 3]{shen2018fractional} and the quadratic transformation from \cite[Corollary 1]{shen2018fractional}, we can rewrite $\widebar{r}_{\mathrm{abs}}$ equivalently as follows
\begin{equation}\notag
\begin{aligned}
    &\widetilde{r}_{\mathrm{abs}}(\bm{\eta},\bm{\zeta},\bm{\gamma},\bm{\alpha})=\frac{1}{K}\left(\sum_{k\in\mathcal{K}}\frac{1}{R_k^{\mathrm{des}}}\log\left(1+\gamma_k\right)-\sum_{k\in\mathcal{K}}\frac{\gamma_k}{R_k^{\mathrm{des}}}\right.\\[-5pt]
    &\left.+\sum_{k\in\mathcal{K}}\left(2\alpha_k\sqrt{(1+\gamma_k)\eta_k}-\alpha_k^2R_k^{\mathrm{des}}(\eta_k+\zeta_k)\right)\right),
\end{aligned}
\end{equation}
where $\bm{\gamma}=[\gamma_1,\cdots,\gamma_K]^T\in \mathbb{R}^K$ and $\bm{\alpha}=[\alpha_1,\cdots,\alpha_K]^T\in\mathbb{R}^K$ denotes two auxiliary vectors, respectively. Note that $\widebar{r}_{\mathrm{abs}}(\bm{\eta},\bm{\zeta},\bm{\gamma},\bm{\alpha})$ is concave with respect to $\bm{\eta},\bm{\zeta}$ for fixed $\bm{\gamma},\bm{\alpha}$. Next, by exploiting the monotonically increasing nature of the $\log(\cdot)$ function, we can recast secrecy constraint C3 as follows:
\begin{equation}\notag
\begin{aligned}
    \Leftrightarrow& \hspace{1mm}\overline{\mbox{C3}}:\max_{\widetilde{\mathbf{H}}_{e,j}[t_0]}\mathrm{Tr}(\widetilde{\mathbf{H}}_{e,j}^H[t_0]\mathbf{V}\widetilde{\mathbf{H}}_{e,j}[t_0]\mathbf{W}_k)\leq \sigma_{e,j}^2(2^{R_{k,j}^S}-1).
\end{aligned}
\end{equation}

Moreover, constraints $\overline{\mbox{C3}}$, C7 and C8 are intractable since they correspond to infinitely many non-convex inequality constraints because of the continuity of the CSI uncertainty set. Therefore, in the following, we recast constraints $\overline{\mbox{C3}}$, C7, and C8 to a finite number of LMI constraints by exploiting the S-Procedure \cite{boyd2004convex}.
We first convert constraint $\overline{\mbox{C3}}$ to an equivalent LMI constraint. By defining $\mathbf{g}_{e,j}=\mathrm{vec}(\widetilde{\mathbf{H}}_{e,j})$, we can rewrite constraint $\overline{\mbox{C3}}$ as follows
\begin{equation}
    \begin{aligned}
        \hspace{-3mm}\overline{\mbox{C3}}\overset{(a)}{\Leftrightarrow}\max_{{\mathbf{g}}_{e,j}}{\mathbf{g}}_{e,j}^H(\mathbf{V}^T\bigotimes\mathbf{W}_k){\mathbf{g}}_{e,j}-(2^{R_{k,j}^S}-1)\sigma_{e,j}^2\leq 0,
    \end{aligned}
\end{equation}
where equality (a) holds due to $\mathrm{Tr}(\mathbf{A}^H\mathbf{BCD})=\mathrm{vec}^H(\mathbf{A})(\mathbf{D}^T\bigotimes\mathbf{B})\mathrm{vec}(\mathbf{C})$. Moreover, let $\widebar{\mathbf{g}}_{e,j}=\mathrm{vec}(\widehat{\widebar{\mathbf{H}}}_{e,j})$ and $\Delta\mathbf{g}_{e,j}[t_0]=\mathrm{vec}(\Delta\widetilde{\mathbf{H}}_{e,j}[t_0])$, then ${\mathbf{g}}_{e,j}[t_0]$ can be equivalently rewritten as ${\mathbf{g}}_{e,j}[t_0]=\widebar{\mathbf{g}}_{e,j}[t_0]+\Delta\mathbf{g}_{e,j}[t_0]$,
where $\|\Delta\mathbf{g}_{e,j}\|_2=\|\Delta\mathbf{H}_{e,j}\|_F=\epsilon_{\mathbf{H}_{e,j}}$. Then, constraint $\overline{\mbox{C3}}$ can be further recast as
\begin{equation}\notag
    \begin{aligned}
        &\max_{{\mathbf{g}}_{e,j}[t_0]}\hspace{-0.5mm}\widebar{\mathbf{g}}_{e,j}^H(\mathbf{V}^T\hspace{-0.5mm}\bigotimes\hspace{-0.5mm}\mathbf{W}_k)\widebar{\mathbf{g}}_{e,j}\hspace{-0.5mm}+\hspace{-0.5mm}\Delta{\mathbf{g}}_{e,j}^H[t_0](\mathbf{V}^T\hspace{-0.5mm}\bigotimes\hspace{-0.5mm}\mathbf{W}_k)\Delta{\mathbf{g}}_{e,j}[t_0]\\
        &+2\mathrm{Re}(\widebar{\mathbf{g}}_{e,j}^H(\mathbf{V}^T\bigotimes\mathbf{W}_k)\Delta{\mathbf{g}}_{e,j}[t_0])-(2^{R_{k,j}^S}-1)\sigma_{e,j}^2\leq 0.
    \end{aligned}
\end{equation}
Next, by exploiting the S-procedure in \cite{boyd2004convex}, we can equivalently convert constraint $\overline{\mbox{C3}}$ as
follows
\begin{equation}
\label{LMI_SINR}
\begin{aligned}
\mathbf{P}_{e,j}-
    \mathbf{G}_{e,j}^H(\mathbf{V}^T\bigotimes\mathbf{W}_k)\mathbf{G}_{e,j}\succeq \mathbf{0}, \forall k,
\end{aligned}
\end{equation}
where 
\begin{equation}\notag
    \begin{aligned}
        \mathbf{P}_{e,j}&=
    \begin{bmatrix}
        q_{e,j}\mathbf{I}_{M(N+1)} & \mathbf{0} \\
        \mathbf{0} & -q_{e,j} \epsilon_{\widetilde{\mathbf{H}}_{e,j}}^2+(2^{R_{k,j}^S}-1)\sigma_{e,j}^2
    \end{bmatrix},
    \end{aligned}
\end{equation}
$\mathbf{G}_{e,j}=[\mathbf{I}_{M(N+1)}\quad \Bar{\mathbf{g}}_{e,j}],$ and $q_{e,j}\geq 0$ is an auxiliary variable. Similarly, by defining $\mathbf{g}_{k}[t_0]=\mathrm{vec}(\widetilde{\mathbf{H}}_{k}[t_0])$, $\widebar{\mathbf{g}}_{k}=\mathrm{vec}(\widetilde{\mathbf{H}}_{k}[0])$, and $\Delta\mathbf{g}_{k}[t_0]=\mathrm{vec}(\Delta\widetilde{\mathbf{H}}_{k}[t_0])$, we can also transform constraints C7 and C8 to a number of LMI constraints as follows
\begin{equation}
    \begin{aligned}
        \mbox{C7}&\Leftrightarrow\mathbf{P}_{S,k}+
    \mathbf{G}_{k}^H(\mathbf{V}^T\bigotimes\mathbf{W}_k)\mathbf{G}_{k}\succeq \mathbf{0}, \forall k\\
            \mbox{C8}&\Leftrightarrow\mathbf{P}_{I,k}-
    \mathbf{G}_{k}^H(\mathbf{V}^T\bigotimes\sum_{k'\in\mathcal{K}\setminus k}\mathbf{W}_{k'})\mathbf{G}_{k}\succeq \mathbf{0}, \forall k,\\
    \end{aligned}
\end{equation}
where 
\begin{equation}
    \begin{aligned}
        \mathbf{P}_{S,k}&=
    \begin{bmatrix}
        q_{S,k}\mathbf{I}_{M(N+1)}\hspace{-8mm} & \mathbf{0} \\
        \hspace{-6mm}\mathbf{0} & \hspace{-8mm}-q_{S,k}\epsilon_{ \widetilde{\mathbf{H}}_{k}}^2-\eta_k
    \end{bmatrix},
    \mathbf{P}_{I,k}=
    \begin{bmatrix}
        q_{I,k}\mathbf{I}_{M(N+1)} \hspace{-8mm}& \mathbf{0} \\
        \hspace{-6mm}\mathbf{0} & \hspace{-8mm}-q_{I,k}\epsilon_{ \widetilde{\mathbf{H}}_{k}}^2-\zeta_k
    \end{bmatrix},\notag
    \end{aligned}
\end{equation}
$\mathbf{G}_{k}=[\mathbf{I}_{M(N+1)}\quad \Bar{\mathbf{g}}_{k}]$, where $q_{S,k}$ and $q_{I,k}$ denote the auxiliary variables for constraints C7 and C8, respectively. Note that reformulated constraints $\overline{\mbox{C3}}$, C7, and C8 are convex with respect to (w.r.t.) $\mathbf{W}_k$, $\forall k$, for a given $\mathbf{V}$ and vice versa. We define vectors $\mathbf{q}_{e}=[{q}_{e,1},\cdots, {q}_{e,J}]$, $\mathbf{q}_{S}=[q_{S,1},\cdots, q_{S,K}]^T$, and $\mathbf{q}_{I,k}=[q_{I},\cdots, q_{I,K}]^T$ to collect the auxiliary variables for constraints $\overline{\mbox{C3}}$, C7, and C8, respectively. Then, the absorption optimization problem can be reformulated as follows:
\begin{eqnarray}
\label{LB_problem_reform}
     &&\hspace*{-6mm}\underset{\substack{\mathbf{v},\mathbf{W}_k,\bm{\eta},\bm{\zeta},\bm{\gamma},\bm{\alpha},\\\mathbf{q}_{e},\mathbf{q}_{S},\mathbf{q}_{I,k}}}{\mino}\hspace*{2mm}  -\widetilde{r}_{\mathrm{abs}}(\bm{\eta},\bm{\zeta},\bm{\gamma},\bm{\alpha})\notag\\[-3pt]
    &&\hspace*{-1mm}\mbox{s.t.}\hspace*{5mm} \mbox{C1},\mbox{C2},\overline{\mbox{C3}},\mbox{C4},\mbox{C5}-\mbox{C8} \hspace*{1mm}.
\end{eqnarray}
Next, we follow the commonly adopted AO principle to obtain a suboptimal solution to problem \eqref{LB_problem_reform}. In particular, the optimization variables in \eqref{LB_problem_reform} are decoupled and optimized alternately until convergence. 
\begin{algorithm}[t]
\caption{Absorption Optimization Algorithm}
\begin{algorithmic}[1]
\small
\STATE Set iteration index $i=0$ and generate a feasible $\mathbf{V}^{(0)}$ and $\mathbf{W}_k^{(0)}, \forall k$. Set convergence tolerances $0<\Delta_{\mathrm{abs}}\ll 1$, $0<\Delta_{\mathrm{SCA}}\ll 1$ and penalty factor $0<\mu\ll 1$.
\REPEAT
\STATE Set $i=i+1$
\STATE Update $\bm{\gamma}^{(i)}$ and $\bm{\alpha}^{(i)}$ based on \eqref{Update_gamma} and \eqref{Update_t}, respectively.
\STATE Solve \eqref{Update_W_problem} for given $\mathbf{v}^{(i-1)}$, $\mathbf{V}^{(i-1)}$, $\bm{\gamma}^{(i)}$, and $\bm{\alpha}^{(i)}$ and update $\mathbf{W}_k^{(i)}$, $\forall k$ as the optimal solution of \eqref{Update_W_problem}.
\STATE Set iteration index $j=0$ and generate a feasible $\mathbf{B}^{(0)}$.
\REPEAT
\STATE Set $j=j+1$
\STATE Solve \eqref{SCA_problem} for a given $\mathbf{B}^{(j-1)}$ and update $\mathbf{B}^{(j)}$ as the optimal solution of \eqref{SCA_problem}.
\UNTIL $\frac{\|\mathbf{B}^{(j-1)}-\mathbf{B}^{(j)}\|_F}{\|\mathbf{B}^{(j-1)}\|_F}\leq \Delta_{\mathrm{SCA}}$
\STATE Update $\mathbf{V}^{(i)}=(\mathbf{B}^{(j)})^H\bm{\theta}\bm{\theta}^H\mathbf{B}^{(j)}$
\UNTIL $\frac{\|\widetilde{r}_{\mathrm{abs}}^{(i-1)}-\widetilde{r}_{\mathrm{abs}}^{(i)}\|_2}{\|\widetilde{r}_{\mathrm{abs}}^{(i-1)}\|_2}\leq \Delta_{\mathrm{abs}}$
\end{algorithmic}
\end{algorithm}
\subsubsection{Update $\bm{\gamma}$ and $\bm{\alpha}$:}In the $i$-th iteration of the proposed AO-based algorithm, we first update the auxiliary vectors $\bm{\alpha}$ and $\bm{\gamma}$. In particular, let $\bm{\eta}^{(i-1)}$ and $\bm{\zeta}^{(i-1)}$ denote the values of $\bm{\eta}$ and $\bm{\zeta}$ obtained in the $(i-1)$-th iteration, respectively. For fixed $\bm{\eta}=\bm{\eta}^{(i-1)}$ and $\bm{\zeta}=\bm{\zeta}^{(i-1)}$, the optimal $\bm{\gamma}^{(i)}=[\gamma_1^{(i)},\cdots, \gamma_K^{(i)}]^T$ can be obtained from the condition $\frac{\partial \widebar{r}_{\mathrm{abs}}(\bm{\eta}^{(i-1)},\bm{\zeta}^{(i-1)},\bm{\gamma})}{\partial \gamma_k}=0$, i.e.,
\begin{equation}
\label{Update_gamma}
    \gamma_k^{(i)}={\eta_k^{(i-1)}}/{(\zeta_k^{(i-1)}+\sigma_k^2)},\, \forall k.
\end{equation}
Then, for given $\bm{\gamma}=\bm{\gamma}^{(i)}$, $\bm{\eta}=\bm{\eta}^{(i-1)}$ and $\bm{\zeta}=\bm{\zeta}^{(i-1)}$, the optimal $\bm{\alpha}$ can be determined via $\frac{\partial \widetilde{r}_{\mathrm{abs}}(\bm{\eta}^{(i-1)},\bm{\zeta}^{(i-1)},\bm{\gamma}^{(i)},\bm{\alpha})}{\partial \alpha_k}=0$, and is given by
\begin{equation}
\label{Update_t}
    \alpha_k^{(i)}={\sqrt{(1+\gamma_k^{(i)})\eta_k^{(i-1)}}}/{(R_k^{\mathrm{des}}(\eta_k^{(i-1)}+\zeta_k^{(i-1)}))}.
\end{equation}
\subsubsection{Update of $\mathbf{W}$:} Let $\mathbf{V}^{(i-1)}$ denote the matrix $\mathbf{V}$ obtained in the $(i-1)$-th iteration. For given $\mathbf{V}=\mathbf{V}^{(i-1)}$, $\bm{\gamma}=\bm{\gamma}^{(i)}$, and $\bm{\alpha}=\bm{\alpha}^{(i)}$, the beamforming optimization problem can be recast as follows,
\begin{eqnarray}
\label{Update_W_problem}
     &&\hspace*{-6mm}\underset{\mathbf{W}_k,\bm{\eta},\bm{\zeta},\mathbf{q}_{e},\mathbf{q}_{S},\mathbf{q}_{I}}{\mino}\hspace*{2mm} -\widetilde{r}_{\mathrm{abs}}(\bm{\eta},\bm{\zeta},\bm{\gamma}^{(i)},\bm{\alpha}^{(i)})\notag\\[-3pt]
    &&\hspace*{-1mm}\mbox{s.t.}\hspace*{5mm} \mbox{C2},\overline{\mbox{C3}},\mbox{C5},\mbox{C7},\mbox{C8}\hspace*{1mm},\mbox{C6:}\hspace*{1mm} \mathrm{rank}(\mathbf{W}_k)=1\hspace*{1mm}.
\end{eqnarray}
Note that the objective function and constraints $\mbox{C2},\overline{\mbox{C3}},\mbox{C5},\mbox{C7},\mbox{C8}$ are convex w.r.t. $\mathbf{W}_k,\bm{\eta}$, and $\bm{\zeta}$ for fixed $\mathbf{V}$, $\bm{\gamma}$, and $\bm{\alpha}$. Moreover, we can relax the rank-one constraint C6 without loss of optimality. The tightness of this relaxation can be proven in a similar manner as in \cite[Appendix]{wu2023globallyoptimal}, which is not shown here due to the page limitation. 
After rank-one relaxation, the optimal solution of the optimization problem in \eqref{Update_W_problem} can be obtained by exploiting standard commercial solvers such as CVX\cite{grant2008cvx}. Here, $\mathbf{W}_k^{(i)}$, $\forall k$, denotes the obtained optimal solution for $\mathbf{W}_k$, which is used in the following update of matrix $\mathbf{V}$.

\subsubsection{Update of $\mathbf{V}$:} To tackle the discrete nature of $\theta_n$, we define phase shift vector $\bm{\theta}=[e^{j\Psi_1},\cdots, e^{j\Psi_L}]^T$. We introduce the binary selection vector for the $n$-th phase shifter, denoted by $\mathbf{b}_n=\big[b_n[1],\cdots,b_n[L]\big]^H,\hspace*{1mm}\forall n$, as $b_n[l]\in\left\{0,\hspace*{1mm}1\right\},\hspace*{1mm}\forall l,n,\hspace*{2mm} \sum_{l=1}^{L}b_n[l]=1,\hspace*{1mm}\forall n$. Then, the coefficient of the $n$-th IRS phase shifter can be expressed as $e^{j\theta_n}=\mathbf{b}_n^H\bm{\theta}$. Here, $b_n[l]=1$ if and only if the $l$-th phase shift value is selected, i.e., $e^{j\theta_n}=e^{j\Psi_l}.$ Hence, $\mathbf{v}$ can be expressed as $\mathbf{v}=\mathbf{B}^H\boldsymbol{\theta}$, where matrix $\mathbf{B}\in\mathbb{C}^{L\times(N+1)}$ is defined as $\mathbf{B}=[\mathbf{b}_1,\cdots,\mathbf{b}_{N+1}]$. Note that $\mathbf{b}_{N+1}^H=[1,0,\cdots,0]$ is a constant binary vector. Therefore, equality constraint C4 can be recast as $\overline{\mbox{C4}}: \mathbf{V}=\mathbf{B}^H\bm{\theta}\bm{\theta}^H\mathbf{B}$,
which is a non-convex quadratic equality constraint. According to \cite[Appendix A]{6698281}, we recast constraint $\overline{\mbox{C4}}$ by exploiting the following lemma:
\begin{lemma}
Based on Schur's complement, equality constraint $\overline{\mbox{C4}}$ is equivalent to the following inequality constraints:
\begin{eqnarray}\notag
\overline{\mathrm{{C4a}}}:\hspace*{-1mm}\label{sdp}
   \begin{bmatrix}
        \mathbf{S} & \hspace*{-4mm}\mathbf{V} &\hspace*{-4mm} \mathbf{B}^H\bm{\theta}\\
        \mathbf{V} & \hspace*{-4mm}\mathbf{T} &\hspace*{-4mm} \mathbf{B}^H\bm{\theta}\\
        \bm{\theta}^H\mathbf{B} & \bm{\theta}^H\mathbf{B} & 1
    \end{bmatrix}\hspace*{-0.5mm}\succeq\hspace*{-0.5mm} \mathbf{0},\,\overline{\mathrm{{C4b}}}:\hspace*{-1mm}\label{DC}
    \mathrm{Tr}\left(\mathbf{S}-\mathbf{B}^H\bm{\theta}\bm{\theta}^H\mathbf{B}\right)\hspace*{-0.5mm}\leq\hspace*{-0.5mm}0,\vspace*{-2mm}
\end{eqnarray}
where $\mathbf{S}\in\mathbb{C}^{(N+1)\times (N+1)}\succeq \mathbf{0}$ and $\mathbf{T}\in\mathbb{C}^{(N+1)\times (N+1)}\succeq \mathbf{0}$, are auxiliary optimization variables.\vspace*{-1mm}
\end{lemma}
Here, thanks to the binary nature of binary selection matrix $\mathbf{B}$, we can equivalently rewrite constraint $\overline{\mbox{C4b}}$ as follows
\begin{equation}
\begin{aligned}
    \widetilde{\mbox{C4b}}:\mathrm{Tr}\left(\mathbf{S}-\mathbf{B}^H\bm{\theta}\bm{\theta}^H\mathbf{B}\right)=\mathrm{Tr}\left(\mathbf{S}\right)-(N+1)\leq 0,
\end{aligned}
\end{equation}
which is a convex affine constraint. For the $i$-th iteration, by fixing $\mathbf{W}_k=\mathbf{W}_k^{(i)}$, $\forall k$, $\bm{\gamma}=\bm{\gamma}^{(i)}$, and $\bm{\alpha}=\bm{\alpha}^{(i)}$, we can recast the optimization problem in \eqref{LB_problem_reform} as follows
 \begin{eqnarray}
\label{Reform_V_problem}
     &&\hspace*{-6mm}\underset{\mathbf{V},\mathbf{S},\mathbf{T},\mathbf{B},\bm{\eta},\bm{\zeta}}{\mino}\hspace*{2mm} -\widetilde{r}_{\mathrm{abs}}(\bm{\eta},\bm{\zeta},\bm{\gamma}^{(i)},\bm{\alpha}^{(i)})\notag\\[-3pt]
    &&\hspace*{-1mm}\mbox{s.t.}\hspace*{5mm} \overline{\mbox{C3}},\overline{\mbox{C4a}},\widetilde{\mbox{C4b}},\mbox{C7},\mbox{C8}\hspace*{1mm}\\[-3pt]
    &&\hspace*{4mm}\hspace*{1mm}\mbox{C1a:}\hspace*{1mm}\sum_{l=1}^{L}b_n[l]=1,\hspace*{1mm}\forall n,\,\mbox{C1b:}\hspace*{1mm}b_n[l]\in\left\{0,\hspace*{1mm}1\right\},\hspace*{1mm}\forall l,\hspace*{1mm}\forall n.\notag\vspace*{-3mm}
\end{eqnarray}
Note that the above optimization problem is non-convex due to the binary constraint C1b. To circumvent this difficulty, we recast C1b in form of the following two inequality constraints
 \begin{eqnarray}
     &&\hspace*{-6mm}\overline{\mbox{C1b}}:\hspace*{0mm} \sum_{n=1}^N\sum_{l=1}^L (b_n[l]-b_n^2[l])\leq 0, \overline{\mbox{C1c}}:\hspace*{0mm} 0\leq b_n[l]\leq 1, \hspace*{1mm}\forall l,\forall n,\notag
\end{eqnarray}
where constraint $\overline{\mbox{C1b}}$ is a difference of convex (DC) function constraint. 
To facilitate low-complexity algorithm design, we resort to the penalty method \cite{ng2015secure} to tackle DC constraint $\overline{\mbox{C1b}}$ and 
rewrite the above problem as follows 
\begin{eqnarray}
\label{Penalty_problem}
    &&\hspace*{-6mm}\underset{\mathbf{V},\mathbf{S},\mathbf{T},\mathbf{B},\bm{\eta},\bm{\zeta}}{\mino}\hspace*{0mm} -\widetilde{r}_{\mathrm{abs}}(\bm{\eta},\bm{\zeta},\bm{\gamma}^{(i)},\bm{\alpha}^{(i)})+\frac{1}{\mu}\sum_{n=1}^N\sum_{l=1}^L (b_n[l]-b_n^2[l])\notag\\[-3pt]
    &&\hspace*{0mm}\mbox{s.t.}\hspace*{8mm}\mbox{C1a}, \overline{\mbox{C1c}},\overline{\mbox{C3}},\overline{\mbox{C4a}},\widetilde{\mbox{C4b}},\mbox{C7},\mbox{C8},
\end{eqnarray}
where $\mu>0$ is a penalty factor for penalizing the violation of constraint $\overline{\mbox{C1b}}$. 
Here, a suboptimal solution of \eqref{Penalty_problem} can be obtained by successive convex approximation (SCA) \cite{le2012exact}. 
In particular, in the $j$-th iteration of the SCA algorithm, we solve the following optimization problem,
    \begin{eqnarray}
\label{SCA_problem}
    &&\hspace*{-6mm}\underset{\mathbf{V},\mathbf{S},\mathbf{T},\mathbf{B},\bm{\eta},\bm{\zeta}}{\mino}\hspace*{2mm} -\widetilde{r}_{\mathrm{abs}}(\bm{\eta},\bm{\zeta},\bm{\gamma}^{(i)},\bm{\alpha}^{(i)})\notag\\[-7pt]
    &&\hspace*{10mm}+\frac{1}{\mu}\sum_{n=1}^N\sum_{l=1}^L (b_n[l]-2b_n^{(j-1)}[l]b_n[l]+(b_n^{(j-1)}[l])^2)\notag\\[-3pt]
    &&\hspace*{0mm}\mbox{s.t.}\hspace*{8mm}\mbox{C1a}, \overline{\mbox{C1c}},\overline{\mbox{C3}},\overline{\mbox{C4a}},\widetilde{\mbox{C4b}},\mbox{C7},\mbox{C8}.
\end{eqnarray}
where $b_n^{(j-1)}[l]$ is the solution for $b_n[l]$ in the $(j-1)$-th iteration of the SCA method. Problem \eqref{SCA_problem} is convex and can be optimally solved by standard convex program solvers such as CVX.
The overall algorithm for the absorption optimization design is summarized in \textbf{Algorithm 1}. In particular, by iteratively updating variables $\mathbf{V}$, $\mathbf{W}_k$, $\bm{\eta},\bm{\zeta},\bm{\gamma},$ and $\bm{\alpha}$, the proposed suboptimal algorithm converges to a stationary point of \eqref{LB_problem_reform} with polynomial time computational complexity \cite{shen2018fractional,yu2021robust}. 

\subsection{Adaptation Optimization Design}

First, we exploit the Lagrangian dual transform and quadratic transform in \cite{shen2018fractional} to reformulate \eqref{Ori_problem_Ada} as
\begin{eqnarray}
\label{Frac_problem_Ada}
     &&\hspace*{-6mm}\underset{\mathbf{W}_k,\hat{\bm{\alpha}},\hat{\bm{\gamma}}}{\mino}\hspace*{2mm}  -\frac{1}{K}\sum_{k\in\mathcal{K}}\left(2\hat{t}_k\sqrt{(1+\hat{\gamma}_k)\mathrm{Tr}(\widetilde{\mathbf{H}}_k[t_n]\mathbf{W}_k)}-\frac{\hat{\gamma}_k}{R_k^{\mathrm{des}}}\right.\notag\\[-4pt]
     &&\hspace*{-6mm}\left.+\frac{1}{R_k^{\mathrm{des}}}\log\left(1+\hat{\gamma}_k\right)-\hat{t}_k^2R_k^{\mathrm{des}}(\sum_{k'}\mathrm{Tr}(\widetilde{\mathbf{H}}_k[t_n]\mathbf{W}_{k'})+\sigma_k^2)\right),\notag\\[-4pt]
    &&\hspace*{-1mm}\mbox{s.t.}\hspace*{5mm} \mbox{C2},\mbox{C3},\mbox{C5},\mbox{C6}.
\end{eqnarray}
The above formulated problem is convex w.r.t. $\mathbf{W}_k$ for fixed $\hat{\bm{\alpha}},\hat{\bm{\gamma}}$, while the optimal solution for $\hat{\bm{\alpha}},\hat{\bm{\gamma}}$ can be obtained in closed-form for fixed $\mathbf{W}_k$. Therefore, we optimize the adaptation performance based on the AO principle. In the $i$-th iteration, we first update the auxiliary vectors $\hat{\bm{\alpha}}$ and $\hat{\bm{\gamma}}$. Let $\widehat{\mathbf{W}}^{(i-1)}_k$ denote the solution for $\mathbf{W}_k$ obtained in the $i$-th iteration. By fixing $\mathbf{W}_k=\widehat{\mathbf{W}}^{(i-1)}_k$, the optimal $\hat{\gamma}_k^{(i)}$ can be obtained by solving $\partial\widetilde{r}_{\mathrm{ada}}(\widehat{\mathbf{W}}^{(i-1)}_k,\hat{\bm{\gamma}})/\partial \hat{\gamma}_k=0$, i.e., $  \hat{\gamma}_k^{(i)}=\frac{\mathrm{Tr}(\widetilde{\mathbf{H}}_k[t_n]\widehat{\mathbf{W}}^{(i-1)}_k)}{\sum_{k'\neq k}\mathrm{Tr}(\widetilde{\mathbf{H}}_k[t_n]\widehat{\mathbf{W}}_{k'}^{(i-1)})+\sigma_k^2}.$
Next, by fixing $\mathbf{W}_k=\widehat{\mathbf{W}}_k^{(i-1)}$ and $\hat{\bm{\gamma}}=\hat{\bm{\gamma}}^{(i)}$, the optimal $\bm{\alpha}^{(i)}$ is obtained according to $    \hat{t}_k^{(i)}=\frac{\sqrt{(1+\hat{\gamma}_k^{(i)})\mathrm{Tr}(\widetilde{\mathbf{H}}_k[t_n]\widehat{\mathbf{W}}^{(i-1)}_k)}}{\sum_{k'\neq k}\mathrm{Tr}(\widetilde{\mathbf{H}}_k[t_n]\widehat{\mathbf{W}}_{k'}^{(i-1)})+\sigma_k^2}.$
Then, after fixing $\hat{\bm{\alpha}}=\hat{\bm{\alpha}}^{(i)}$, $\hat{\bm{\gamma}}=\hat{\bm{\gamma}}^{(i)}$ and relaxing rank-one constraint C7, the adaptation optimization problem in \eqref{Frac_problem_Ada} is convex w.r.t. $\mathbf{W}_k$, which can be optimally solved.
Here, the tightness of the relaxation of constraint C7 can be proven in a similar manner as in \cite[Appendix]{wu2023globallyoptimal}. The overall procedure of the proposed adaptation optimization algorithm is summarized in \textbf{Algorithm 2}. Based on \cite{shen2018fractional}, the proposed algorithm is guaranteed to obtain a stationary point of problem \eqref{Ori_problem_Ada} in polynomial time.
\begin{algorithm}[t]
\caption{Adaptation Optimization Algorithm}
\begin{algorithmic}[1]
\small
\STATE Set iteration index $i=0$ and generate a feasible $\widehat{\mathbf{W}}_k^{(0)}, \forall k$. Set $0<\Delta_{\mathrm{ada}}\ll 1$. Initialize $\hat{\gamma}_k^{(0)}=\frac{\mathrm{Tr}(\widetilde{\mathbf{H}}_k[t_n]\widehat{\mathbf{W}}^{(0)}_k)}{\sum_{k'\neq k}\mathrm{Tr}(\widetilde{\mathbf{H}}_k[t_n]\widehat{\mathbf{W}}_{k'}^{(0)})+\sigma_k^2}$
\REPEAT
\STATE Set $i=i+1$
\STATE Update $\hat{\bm{\gamma}}^{(i)}$ and $\hat{\bm{\alpha}}^{(i)}$.
\STATE Solve \eqref{Frac_problem_Ada} by relaxing C6 for given $\hat{\bm{\gamma}}^{(i)}$, $\hat{\bm{\alpha}}^{(i)}$ and update $\widehat{\mathbf{W}}_k^{(i)}$
\UNTIL $\frac{\|\hat{\bm{\gamma}}^{(i-1)}-\hat{\bm{\gamma}}^{(i)}\|_2}{\|\hat{\bm{\gamma}}^{(i-1)}\|_2}\leq \Delta_{\mathrm{ada}}$
\end{algorithmic}
\end{algorithm}
\section{Numerical Results}
In this section, we consider a system with a BS equipped with $M=6$ transmit antennas to provide communication services to $K=4$ users. The IRS is deployed $D=40$ m from the BS, while the users and $J=2$ Eves are located on circles having their origins at the IRS and radii of $r=5$ m and $r_e=20$ m, respectively. The channel matrix $\mathbf{F}[0]$ between the BS and the IRS in the initial time slot is modeled as Rician fading
\begin{equation}
    \hspace{-2mm}\mathbf{F}[0]=\sqrt{L_0D^{-\alpha_{\mathrm{BI}}}}\left(\sqrt{\frac{\beta_{\mathrm{BI}}}{1+\beta_{\mathrm{BI}}}}\mathbf{F}_L+\sqrt{\frac{1}{1+\beta_{\mathrm{BI}}}}\mathbf{F}_N\right),
\end{equation}
where $L_0$ is the path loss at reference distance $d_0=1$ m. $\alpha_{\mathrm{BI}}=2.2$ and $\beta_{\mathrm{BI}}=3$ denote the path loss exponent and the Rician factor, respectively \cite{yu2021robust}. Matrices $\mathbf{F}_L$ and $\mathbf{F}_N$ are the line-of-sight (LoS) and non-LoS (NLoS) components, respectively. Here, the LoS matrix $\mathbf{F}_L$ is the outer product of the receive and transmit array response vectors, while the entries of NLoS matrix $\mathbf{F}_N$ are modeled as independent Rayleigh-distributed random variables. We generate the other channels in the initial time slot in the system in a similar manner as $\mathbf{F}[0]$. The path loss exponents for the BS-user, BS-Eve, IRS-user, and IRS-Eve links are set to $\alpha_{\mathrm{BU}}=\alpha_{\mathrm{BE}}=3.6$ and $\alpha_{\mathrm{IU}}=\alpha_{\mathrm{IE}}=2.2$, respectively. Moreover, the corresponding Rician factors are given by $\beta_{\mathrm{BU}}=\beta_{\mathrm{BE}}=1$ and $\beta_{\mathrm{IU}}=\beta_{\mathrm{IE}}=3$, respectively. The noise variances are set to $\sigma_k^2=\sigma_{e,j}^2=-90$ dBm, $\forall k \in\mathcal{K}$, $\forall j \in\mathcal{J}$. The transmit power $P_{\mathrm{max}}$ is set to $30$ dBm. The maximal tolerable information leakage to the Eves and the desired user rate are set to $R_{k,j}^S=R^S=0.5$ bits/s/Hz and $R_k^{\mathrm{des}}=R^{\mathrm{des}}=3$ bits/s/Hz, respectively. Here, we set the normalized maximal channel variation of the overall user channel in the considered time frame to $\kappa_k=\epsilon_{ \widetilde{\mathbf{H}}_{k}}/\|\widetilde{\mathbf{H}}_k[0]\|_F=0.15$, $\forall k$. Moreover, the normalized maximal channel estimation error of the overall Eve channels is set to $\kappa_{e,j}=\epsilon_{\widetilde{\mathbf{H}}_{e,j}}/\|\widehat{\widebar{\mathbf{H}}}_{e,j}\|_F=0.2$, $\forall j$. To evaluate the resilience performance, we adopted the system failure model in \cite{weinberger2023ris}. In particular, the user channels vary with maximal norm value, i.e., $\|\Delta\widetilde{\mathbf{H}}_k[t_0]\|_F=\epsilon_{ \widetilde{\mathbf{H}}_{k}}$, $\forall k$, in time slot $t_0$ to model a system failure event. For simplicity, the channels before and after system failure are assumed to be static, i.e., $\widetilde{\mathbf{H}}_k[t_s]=\widetilde{\mathbf{H}}_k[0]$, $\forall t_s=\{1,\cdots, t_0-1\}$ and $\widetilde{\mathbf{h}}_k[t_s]=\widehat{\mathbf{h}}_k[t_0]$, $\forall t_s=\{t_0,\cdots,T_s\}$. The considered failure model is sufficient to show the effectiveness and importance of resilience for the considered system. However, more sophisticated failure models will be considered in future works. 
For comparison, we consider two baseline schemes. For \textbf{Baseline Scheme 1}, the IRS phase shift matrix $\bm{\Phi}_{\mathrm{rdn}}$ is randomly generated, and no resilience procedure is implemented. The beamforming matrix $\mathbf{W}$ is designed by \textbf{Algorithm 2} and both $\mathbf{w}_k$ and $\bm{\Phi}_{\mathrm{rdn}}$ are kept fixed for all time slots. Furthermore, we consider a naive scheme without absorption optimization as \textbf{Baseline Scheme 2}. Here, the BS beamforming matrix and IRS phase shifts are initialized by solving problem \eqref{Ori_problem} with $\kappa_k=0$ based on \textbf{Algorithm 1}. After a system failure is detected, the resilient procedure in \textbf{Algorithm 2} is activated to recover performance.

\begin{figure}[htbp]\vspace*{-6mm}
	\centering
    \begin{minipage}{1.7 in}
                 \centering
         \includegraphics[width=1.55in]{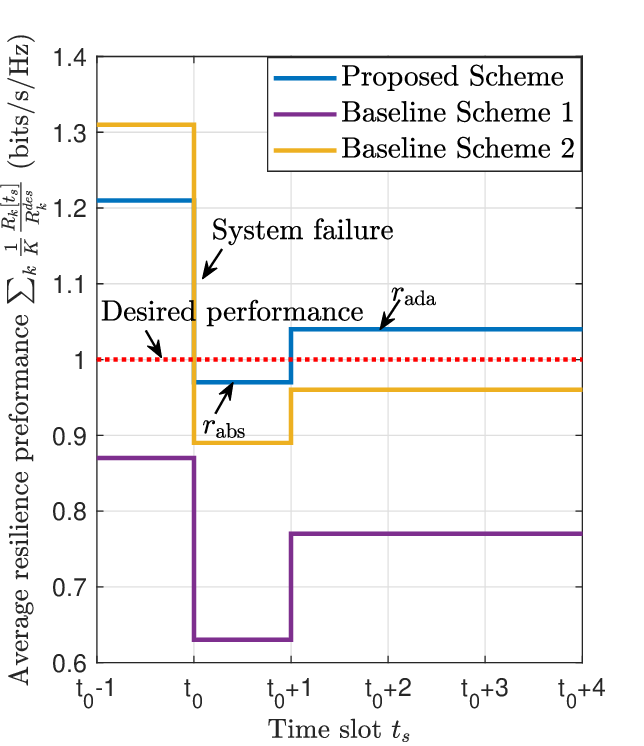}     \label{fig::Behavior}
         \vspace{-0mm}\caption{Resilient behavior of different schemes. The system parameters are set as $N=48$ and $J=2$.}
        
    \end{minipage}
    \begin{minipage}{1.7 in}\vspace{0mm}
                 \centering
         \includegraphics[width=1.55in]{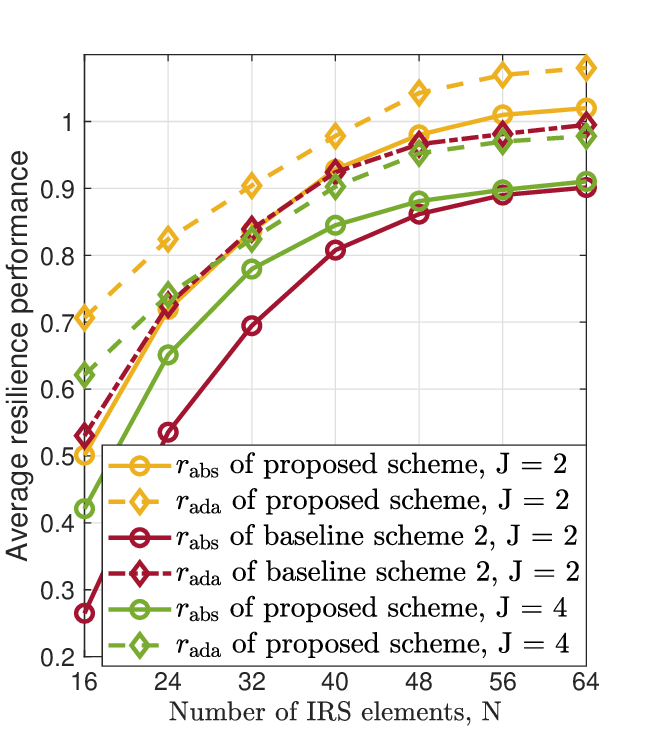}     \label{fig::N_imperfect1}
         \vspace{-0mm}\caption{Absorption and adaptation performance vs. the number of IRS elements for different number of Eves.}
    \end{minipage}
\end{figure}\vspace{-3mm}
In Fig. 3, we investigate the resilience behavior of the proposed design and the baseline schemes. It is observed that the proposed scheme outperforms Baseline Scheme 1 before and after system failure. Since Baseline Scheme 1 employs random IRS phase shifts, the IRS cannot generate a suitable beam to mitigate the system failure, leading to unsatisfactory performance. In contrast, the proposed design achieves the best average user rate after system failure, which confirms its excellent absorption performance. Moreover, the proposed scheme exhibits a better recovery performance compared to Baseline Scheme 2. In particular, since Baseline Scheme 2 employs a non-robust IRS phase shift design, it cannot form a reliable IRS-assisted link after system failure, leading to a $10\%$ rate loss compared to the proposed algorithm.  

Fig. 4 illustrates the resilience performance of the proposed scheme and Baseline Scheme 2 for different system settings. As can be observed, as the number of IRS elements increases, the absorption and adaptation performances of both the proposed scheme and Baseline Scheme 2 are enhanced. In particular, increasing the number of IRS elements $N$ introduces extra degrees of freedom for system design, leading to improved robustness and recovery performance. Moreover, the proposed scheme outperforms Baseline Scheme 2 w.r.t. both the absorption and adaptation metrics. 
In particular, 
the proposed scheme with $J=2$ ensures almost failure-free transmission, i.e., $r_{\mathrm{abs}}\approx 1$, for a moderate IRS size, i.e., $N=48$. 
Moreover, for $N=64$, the proposed scheme can ensure failure-free transmission even for $J=4$ Eves, highlighting its ability to maintain resilience and secrecy in challenging scenarios.

\section{Conclusion}
This paper presents a resilience strategy for an IRS-assisted secure multiuser MISO system. We propose a two-timescale transmission protocol involving an initialization phase design and a recovery mechanism to enhance the absorption and adaptation performance of the system, respectively. Specifically, in the initialization phase, the long-term IRS phase shifts and BS beamforming vectors are computed to guarantee a worst-case achievable rate of the users while limiting the maximum information leakage to Eves based on an initial estimate of the CSI. Meanwhile, short-term BS beamforming vectors are configured after system failure to facilitate fast recovery. Our numerical results reveal that our proposed algorithms enhance the system's resilience significantly compared to two baseline methods. Moreover, for systems with a sufficiently large IRS, the proposed schemes enable failure-free transmission in the presence of strong channel variations over a long period of time. In future work, we plan to exploit AN to further improve the secrecy performance of the considered system.
\bibliographystyle{IEEEtran}
\vspace{-0mm}\bibliography{Reference_List}
\end{document}